\documentclass[aps,prl,showpacs,twocolumn,superscriptaddress]{revtex4}
\usepackage{bm,color,amsmath,amssymb,mathrsfs,latexsym,graphicx,psfrag}

\newcommand{\bs}[1]{\boldsymbol{#1}}

\def\ie{{\it i.e.},\ }

\begin{document}
\title{Doping-dependent pairing symmetry in the Iron-Pnictides}
\author{Ronny Thomale} \affiliation{Institut f\"ur Theorie der
  Kondensierten Materie, Universit\"at Karlsruhe, D 76128 Karlsruhe}
\author{Christian Platt} \affiliation{Theoretical Physics, University of W\"urzburg, D-97074 W\"urzburg}\author{Jiangping Hu} \affiliation{Department of
  Physics, Purdue University, West Lafayette, Indiana 47907}
\author{Carsten Honerkamp} \affiliation{Theoretical Physics, University of W\"urzburg, D-97074 W\"urzburg} \author{B. Andrei Bernevig} \affiliation{Department of Physics, Princeton University, Princeton,
  NJ 08544}


\begin{abstract}
  We use the functional renormalization group method to analyze the
  phase diagram of a 4-band model for the iron-pnictides subject to
  band interactions with certain $A_{1g}$ momentum dependence. We
  determine the parameter regimes where an extended $s$-wave pairing
  instability with and without nodes emerges.  On the electron-doped
  side, the parameter regime in which a nodal gap appears is found to
  be much narrower than recently predicted
  in~\cite{chubukov-09cm09035547}.  On the hole-doped side, the
  extended $s$-wave pairing never becomes nodal: above a critical
  strength of the intra-band repulsion, the system favors an exotic
  extended $d$-wave instability on the enlarged hole pockets at much
  lower $T_c$. At half filling, we find that a strong momentum
  dependence of inter-band pair hopping yields an extended s-wave
  instability instead of spin-density wave (SDW) ordering. These
  results demonstrate that an interaction anisotropy around the Fermi
  surfaces generally leads to a pronounced sensitivity of the pairing
  state on the system parameters.

\end{abstract}
\date{\today}


\maketitle

{\it Introduction---} The discovery of high-temperature
superconductivity in iron arsenide and related compounds at the
beginning of 2008~\cite{kamihara-08jacs3296} has triggered an enormous
interest in condensed matter physics. This new class of materials
exhibits transition temperatures $T_c$ beyond the conventional BCS
regime upon electron~\cite{chen-08prl247002} and hole
doping~\cite{rotter-08prl107006} of a collinear antiferromagnetic
parent state, with $T_c$'s extending up to 56
K~\cite{chen-08n761,wang-08epl67006}, thereby breaking the cuprate
monopoly on high-temperature superconductivity ~\cite{berg-09cm0905}.
Present experimental evidence accompanied by theoretical modelling
suggest that the pairing in the iron-pnictides is different from the
$d$-wave in cuprates.  This could be due to both the different
strength of the electron-electron interactions in the two materials,
which results in an itinerant antiferromagnet in the parent pnictide
compounds, and to the topology of the Fermi surfaces (FS) with complex
multi-band
character~\cite{kuroki-08prl087004,cao-08prb220506}. However, very
similar to the cuprates is the belief that the magnetism of the parent
state crucially influences the pairing symmetry of the doped system.

Various approaches have been pursued to investigate the pairing
symmetry in the iron-pnictides.  By now, after a short period of
analysis providing a wide-spread range of possible
pairings~\cite{si-08prl076401,lee-08prb144517,dai-08prl057008}, the
general theoretical view started to converge to an extended $s$-wave
order parameter (denoted $s^\pm $ or $s_{x^2 y^2}$) that takes
opposite signs on the electron and hole pockets along the multi-band
FS, which is consistent with some experimental data and also has broad
theoretical
support~\cite{mazin-08prl057003,kuroki-08prl087004,seo-08prl206404,wang-09prl047005,parish-08prb144514,
  cvetkovic-09epl37002,seo-09prb235207}.  However, experimentally, there is still no
broad consensus about the nature of the order parameter. While most
experiments can be explained in the framework of an $s^{\pm}$
gap~\cite{parish-08prb144514}, several facts, such as the $T^3$
dependence of the NMR penetration depth over a significant temperature
range, as well as the linear penetration depth in LaOFeP remain
unsettled. Other experiments such as penetration depth on the 1111 and
122 compounds, as well as thermal conductivity can be explained by an
$s^{\pm}$ order parameter but with large gap anisotropy. Gaps with
significant momentum dependence, but no nodes, are reported
experimentally in~\cite{kondo-08prl147003}. In contrast, ARPES data
reveals very isotropic nodeless gaps on the hole Fermi surfaces
\cite{wray-08prb184508,koitzsch-08prb180506,liu-08prb184514}, of
magnitudes matching a strong-coupling form $\Delta(k) = \Delta_0
\cos(k_x) \cdot \cos(k_y)$ \cite{seo-08prl206404} in the unfolded
Brillouin zone.  Non-nodal gaps have been found in several analytical
and numerical theoretical approaches to the
pnictides~\cite{platt-09cm09031963,wang-09prl047005,korshunov-08prb140509,chubukov-08prb134512,stanev-08prb184509}.
Functional renormalization group (fRG) studies \cite{wang-09prl047005} with
orbital interactions reveal the presence of largely anisotropic,
almost nodal, gaps.  A recent RPA analysis suggests the existence of a
nodal extended $s$-wave state within the 5-orbital Hubbard
model~\cite{graser-09njp025016}. It was also recently predicted that
the momentum dependence of the interaction along the electron pockets
may result in the development of gap anisotropy and possibly nodes on
top of the constant $s^{\pm}$
signal~\cite{chubukov-09cm09024188,maier-09cm09035216}. In the strong
coupling mean-field picture \cite{seo-08prl206404,parish-08prb144514},
the gap anisotropy is \emph{doping dependent}: the gap has a form
$\cos(k_x) \cdot \cos(k_y)$ which becomes more anisotropic as the
doping is increased.  In a weak coupling expansion of FS interactions,
the gap anisotropy can arise from the presence of an $A_{1g}$ term
$\cos(k_x) + \cos(k_y)$ (which does not break the crystal symmetry but
can create nodes on the $(\pi,0)$ and $(0,\pi)$ electron surfaces) in
the band interactions \cite{chubukov-09cm09024188}.

In this Letter, we study the influence of interaction anisotropy on
the superconducting instability by
(fRG)~\cite{shankar94rmp129,honerkamp-01prb035109,halboth-00prb7364}
methods.  Our aim is to investigate the possibility that details of
the material and the theoretical model might play a decisive role in
determining the pairing symmetry.  In order to have a simple control
and minimal model of the interaction anisotropy that allows to
generate gap nodes, we do not consider interactions defined in an
orbital representation, but use constant band interactions that only
depend on whether the external legs are on hole- or
electron-pockets. To these we add an $A_{1g}$ momentum dependence in
the inter-band pair hopping interactions $g_3$ with zero total
momentum, as suggested recently in \cite{chubukov-09cm09024188}.  For
the electron doped regime, we mostly find non-nodal $s^{\pm}$, where
the scale of the gap anisotropy increases with enhanced intra-band
repulsion $g_4$. For highly dominant $g_4$, we indeed find a small
parameter regime of nodal $s^{\pm}$. On the hole-doped side, we first
find a rather isotropic $s^{\pm}$ signal where the scale of anisotropy
does not decisively depend on $g_4$. Upon further increasing
$g_4$ and hole doping, a phase transition occurs, where both the
electron and the hole pockets become nodal: the hole pockets develop a
$d$-wave intra-band cooper pairing. This clearly demonstrates the
possibility of a doping-dependent gap function.  Furthermore, we find
that dominant momentum dependence of $g_3$ can increase $T_c$ of
$s^{\pm}$-wave pairing at half filling such that it even exceeds SDW
as the previously leading instability. Recent experimental evidence
for a superconducting parent compound leaves open the possibility to
have a leading superconducting instability even at half
filling~\cite{zhu-09cm09041732}.

 \begin{figure}[t]
  \begin{minipage}[c]{1.0\linewidth}
    \includegraphics[width=\linewidth]{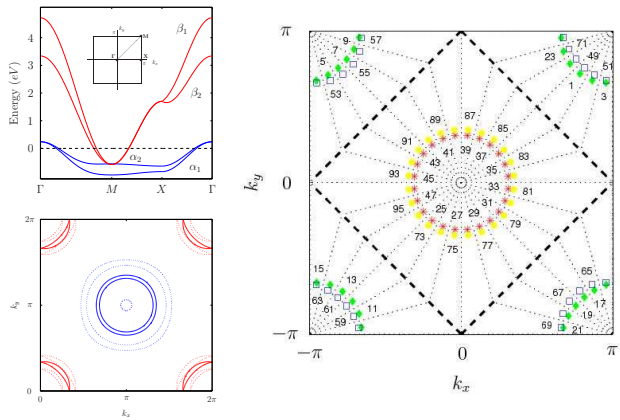}
  \end{minipage}
  \caption{(Color online) (a) The band structure of the four band
    model. Electron pocket bands are denoted by $\beta_{1,2}$ and hole
    pocket bands by $\alpha_{1,2}$. The dashed line shows the Fermi
    level at half filling. (b) Change of Fermi surfaces upon
    doping. Bold lines are FS's at half filling. Upon electron (hole)
    doping ($x=-.15$ and $x=.15$ shown as dashed lines), the electron
    (hole) pockets grow while the hole (electron) pockets shrink. (c)
    Illustration of the patching segments in the Brillouin zone. The
    patches $1,\dots,24$ and $49,\dots, 72$ belong to the electron
    pockets situated at the $M$ point, $25,\dots,48$ and $73,\dots,96$
    to the hole pockets situated at the $\Gamma$ point.}
  \label{fig:band}
\vspace{-10pt}
\end{figure}

{\it 4-band model---} As a tight-binding model, we constrain ourselves
to the 4-band model given by Korshunov and
Eremin~\cite{korshunov-08prb140509} containing the decisive weight of
the density of states in the vicinity of the FS's (see
Fig~\ref{fig:band}). In the folded BZ with two Fe ions per unit cell,
the kinetic Hamiltonian reads:

\begin{equation}
H_0=-\sum_{\bs{k},i,\sigma}\epsilon^i n_{\bs{k} i \sigma} -\sum_{\bs{k},i,\sigma} t_{\bs{k}}^i d_{\bs{k} i \sigma}^\dagger d_{\bs{k} i \sigma}^{\phantom{\dagger}},
\end{equation}
\noindent
where $i$ denotes the band index $i=\alpha_1, \alpha_2, \beta_1,
\beta_2$, and the $\epsilon^i$'s are the on-site energies. The hopping
dispersion parameters for the $\alpha$ bands around the $\Gamma$ point
are given by $t_{\bs{k}}^{\alpha}=t_1^{\alpha}(\cos k_x+\cos
k_y)+t_2^{\alpha} \cos k_x \cos k_y $, whereas parameters for the
$\beta$ bands around the $M$ point are given by
$t_{\bs{k}}^{\beta}=t_1^{\beta}(\cos k_x+\cos k_y)+t_2^{\beta} \cos
k_x/2 \cos k_y/2 $. In units of eV and grouped by
$(\epsilon^i,t^i_1,t^i_2)$, the parameters take on the values
$(-0.60,0.30,0.24)$ for $\alpha_1$, $(-0.40,0.20,0.24)$ for
$\alpha_2$, $(1.70,1.14,0.74)$ for $\beta_1$, and $(1.70,1.14,-0.64)$
for $\beta_2$. The chemical potential can be adjusted such that two
hole pockets emerge at the $\Gamma$ point, whereas two electron
pockets emerge at the $M$ point. At half filling, \ie $\mu=0$, hole
and electron pockets are almost perfectly nested.

{\it Band interactions---} We distinguish four types of band
interactions. First, there are two types of inter-band interaction
vertices depending on whether momentum from ingoing and outgoing
propagators is transferred within the same band or between the bands:

\begin{eqnarray}
H_I^{\text{inter}}=\sum_{\substack{\bs{k},\bs{k'},\bs{q}}}\hspace{-18pt} && g_1 \;( d^{\dagger}_{\bs{k}+\bs{q} \alpha \sigma} d^{\dagger}_{\bs{k'}-\bs{q} \beta \sigma'} d^{\phantom{\dagger}}_{\bs{k'} \alpha \sigma'}d^{\phantom{\dagger}}_{\bs{k} \beta \sigma} + \text{h.c.}) \nonumber \\ &+& g_2 \;( d^{\dagger}_{\bs{k}+\bs{q} \alpha \sigma} d^{\dagger}_{\bs{k'}-\bs{q} \beta \sigma'} d^{\phantom{\dagger}}_{\bs{k'} \beta \sigma'}d^{\phantom{\dagger}}_{\bs{k} \alpha \sigma}+ \text{h.c.}),\nonumber\\
\label{u1u2}
\end{eqnarray}
with implicit sums over $\sigma, \sigma'$, and the band indices
$\alpha$ and $\beta$ extending over $\alpha_{1,2}$ and $\beta_{1,2}$,
respectively (this convention is kept throughout the article). While
$g_1$ turns out to be rather unimportant with respect to the analysis
of the leading instabilities driven by band interactions, $g_2$ is
necessary (but not sufficient) to drive the SDW instability in the
$(\pi, \pi)$ channel between electron and hole pockets in the folded
BZ. Furthermore, we consider the inter-band pair hopping interaction
$g_3$:

\begin{equation}
H_I^{\text{pair}}=\sum_{\substack{\bs{k},\bs{k'},\bs{q}}}g_3^b \;( d^{\dagger}_{\bs{k}+\bs{q} \alpha \sigma} d^{\dagger}_{\bs{k'}-\bs{q} \alpha \sigma'} d^{\phantom{\dagger}}_{\bs{k'} \beta \sigma'}d^{\phantom{\dagger}}_{\bs{k}  \beta \sigma}+ \text{h.c.}) ,
\end{equation}
where, as one central point of our analysis, we include a momentum
dependence of the $A_{1g}$-projected pair hopping amplitude for the
zero momentum cooper channel~\cite{chubukov-09cm09035547}:
\begin{eqnarray}
g_3^b\vert_{\bs{k'}=-\bs{k}}&=& g_3 \left( 1+ b (\cos
 k_x + \cos k_y) \right)\nonumber \\
&& \hspace{-5pt} \cdot \, \left( 1+ b (\cos
 (k_x+q_x) + \cos (k_y+q_y) ) \right),
\label{eq:fu3}
\end{eqnarray}
and constant $g_3$ otherwise. $b$, the anisotropy scale, gives the
relative scale of momentum dependence. To make connection
to~\cite{chubukov-09cm09035547}, $g_3$ defined in~\eqref{eq:fu3} is
given in terms of the unfolded (u) momenta, which relate to the folded
(f) ones by $k_{\text{u} \; x,y}=(k_{\text{f}\; x} \pm k_{\text{f}\;
  y})/2$.  Finally, there is the intra-band pair interaction

\begin{equation}
H_I^{\text{intra}}=\sum_{\substack{\bs{k},\bs{k'},\bs{q}}}g_4 \; d^{\dagger}_{\bs{k}+\bs{q}  i \sigma} d^{\dagger}_{\bs{k'}-\bs{q}  i \sigma'} d^{\phantom{\dagger}}_{\bs{k'}  i \sigma'}d^{\phantom{\dagger}}_{\bs{k}  i \sigma},
\end{equation}
where $i$ extends over all band indices. For $H_I^{\text{pair}}$ in
the total zero momentum Cooper channel, one Cooper pair belongs to the
electron pockets and the other one to the hole pockets, rendering
$g_3$ to significantly deviate depending on the momenta along the
electron FS, but gives only a constant value $g_3 \approx 1+2b$ on the
hole pockets.

{\it fRG method---} We use the fRG method to study the flow of the
two-particle coupling function
$V^{\Lambda}(\bs{k}_1,a,\bs{k}_2,b,\bs{k}_3,c,\bs{k}_4,d)$, where
$\bs{k}_{1,2}$ ($\bs{k}_{3,4}$) denote the ingoing (outgoing)
particles, $a,\dots,d$ are the different band indices, and $\Lambda$
is the energy cutoff above which the high energy contributions are
integrated out and incorporated into the effective coupling
function. Details on the implementation for the multi-band case of
pnictides can be reviewed
in~\cite{platt-09cm09031963,wang-09prl047005}. While $\bs{k}_4$ is
implicitly given by momentum conservation, the spin convention is
chosen such that $\bs{k}_1$ ($\bs{k}_2$) and $\bs{k}_3$ ($\bs{k}_4$)
have the same spin.
As usual, we omit the frequency dependence of the vertex, and assume
that the relevant processes are located in the close vicinity of the
FS. This, together with the neglect of self-energy corrections, is
tantamount to a weak-coupling approach initially, although
interactions strengths grow large under RG flow. To solve the RG
equations numerically, the momenta in the BZ are discretized as shown
in Fig.~\ref{fig:band}.  Each momentum is confined within two
neighboring dotted lines running from $(\pi, \pi)$ to one of the
corner points $(\pi \pm \pi, \pi \pm \pi)$. For a given continuous
momentum $\bs{k}$ and band index $i$, its value is effectively
projected onto the Fermi momentum value of the $i$th band in the angle
section where $\bs{k}$ is situated.  Testing discretizations of $48$,
$64$, $96$, $128$, and $256$ patches, we find that $N=96$ both
provides sufficient resolution and suitable computation time
performance for our studies. Essentially, the RG flow in terms of the
cutoff parameter $\Lambda$ starts at cutoff energy scales of the order
of the bandwidth, and successively decreases towards the FS. In each
differential step, both particle-particle and particle-hole
contributions influence the evolution of the coupling function.  Quite
generically, these flows lead to strong coupling, i.e. one or more
channels in the coupling function flow to large absolute values at a
critical scale $\Lambda_c$. Comparing the growth of the various
pairing and density-wave channels allows one to compare between
leading and subleading instabilities. For standard cases, the critical
scale $\Lambda_c$ provides a reasonable estimate of the actual critical
temperature $T_c$ for phase transitions when long-range ordering is
possible.

 \begin{figure}[t]

  \begin{minipage}[c]{0.95\linewidth}
    \includegraphics[width=\linewidth]{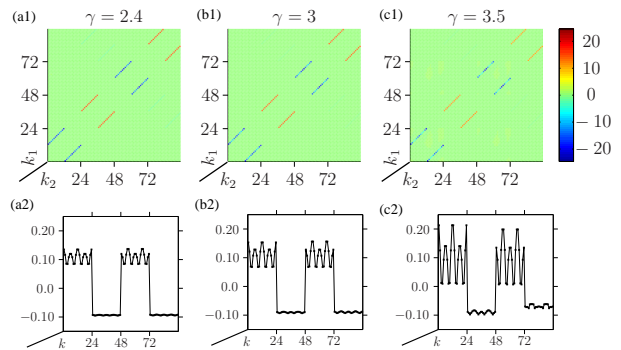}
  \end{minipage}
  \caption{(Color online) (a1)-(c1) Plot of the vertex function
    vs. incoming momenta for $g_1=0.1$,$g_2=0.25$,$g_3=0.3$,$g_4=\gamma
    g_3$, $b=1.1$, and $\gamma$ varied as $2.4$, $3$, and $3.5$ from
    left to right. $k_4$ is fixed by total momentum, $k_3$ is chosen
    to reside on patch sector 55. Upon increase of $\gamma$, one
    clearly observes the nodal structure on the electron pockets to
    become more pronounced, while the hole pockets show a
    homogeneously diverging vertex without anisotropy. (a2)-(c2) Plot
    of the associated superconducting form factor. While the signal
    along the hole pockets is constant, the nodal peaks successively
    develop on the electron pockets.}
  \label{fig:gamma}
\vspace{-10pt}
\end{figure}

\begin{figure}[t]
  \begin{minipage}[c]{0.95\linewidth}
    \includegraphics[width=\linewidth]{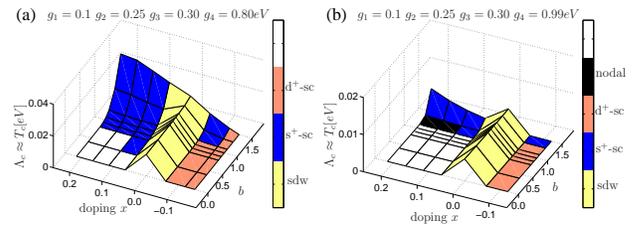}
  \end{minipage}
  \caption{(Color online) Phase diagrams for very strong intra-band
    repulsion $g_4=\gamma g_3$.  $g_1=0.1, g_2=0.25, g_3=0.3,
    g_4=\gamma g_3=0.80$ ($g_4=0.99$) for (a) and (b),
    respectively. (a) Narrow regime around half filling with SDW
    leading instability. $T_c$ is comparably small for the different
    phases, and, for small $b$ and electron doping, undergoes our
    maximum resolution scale denoted by white parcels. The extended
    $d$ wave regime on the hole doped side increases to larger $b$. (b) A small
    parameter window with nodal $s^{\pm}$ and small $T_c$ denoted by
    black is found for electron doping.}
  \label{fig:p2}
\vspace{-10pt}
\end{figure}

{\it Electron doping---}
First we want to specify the parameter window where nodal and
non-nodal $s^{\pm}$ appears for electron doping ($x=0.2$). We consider
the case of $g_1=0.1$ and $g_2=0.25$, and choose $g_3=0.3$ to be of
the order of $g_2$ (throughout the article, the interaction couplings
are given in units of $\text{eV}$ where the total bandwidth is $\sim
6eV$). The $g_3$ anisotropy scale $b$ and $\gamma=g_4/g_3$ span the
relevant parameter space of interactions and allow for nodal gaps
according to Ref. \cite{chubukov-09cm09035547}.  For electron doping,
the pockets nesting is lifted and the SDW instability is
suppressed. For $b \ll 1$, one finds a constant $s^\pm$ instability,
as observed in~\cite{platt-09cm09031963}, whose critical divergence
scale $T_c$ decreases with increasing $\gamma$. The $s^{\pm}$
instability manifests itself as a Cooper instability corresponding to
divergent vertex couplings for $\bs{k}_1=-\bs{k}_2$, with a sign
change of the vertex going from electron to hole pockets, as it can be
seen both in the vertex plot and the form factors as shown in
Fig.~\ref{fig:gamma}. Upon increasing $b$, the critical divergence
scale increases considerably and helps to counteract the intra-band
repulsion $g_4$. In addition, a gap variation starts to emerge on the
electron pockets, while the hole pockets remain unchanged. In
particular, the gap anisotropy increases upon increasing $\gamma$,
which shows that the system favors a nodal variation to compensate the
increasing intra-band repulsion $g_4$, as predicted
in~\cite{chubukov-09cm09035547}. However, for $\gamma \lesssim 3$,
this anisotropy never becomes comparable to the constant gap scale on
the electron pockets, i.e. the nodes do not completely develop. Upon
increasing $\gamma>3$, the gap variation on the electron pockets gets
more pronounced, to finally yield a nodal extended $s$-wave
instability (Fig.~\ref{fig:gamma}c). However, for increasing $b$ even
further at constant $\gamma$, the value of $g_3 \approx 1+2b$ on the
hole pockets is considerably enhanced and leads to a reduction of gap
anisotropy developing on the electron pockets.  Hence, while the
trends pointed out in Ref. \cite{chubukov-09cm09035547} are clearly
visible, we only observe a clean nodal $s^{\pm}$ instability with
sign-changes around the electron pockets in a comparably narrow window
of intermediate $b$ and high $\gamma$ (see Fig.~\ref{fig:gamma} and
Fig.~\ref{fig:p2}). In our understanding, this difference occurs due to
the renormalization of the interactions at higher energy scales before
the pairing instability sets in. This effect is mentioned but not
explicitly taken into account in \cite{chubukov-09cm09035547}, and
hence the parameter ranges for the nodal state differ. Concluding this
part, we note that on the electron-doped side, the fully-gapped
$s\pm$-state is rather stable. It is well possible that fine-tuning of
the band structure like a strengthening of the scattering between the
electron pockets can enhance the anisotropy again, but primary
instability favors an isotropic gap.

{\it Hole Doping---}
\begin{figure}[t]

  \begin{minipage}[c]{0.95\linewidth}
    \includegraphics[width=\linewidth]{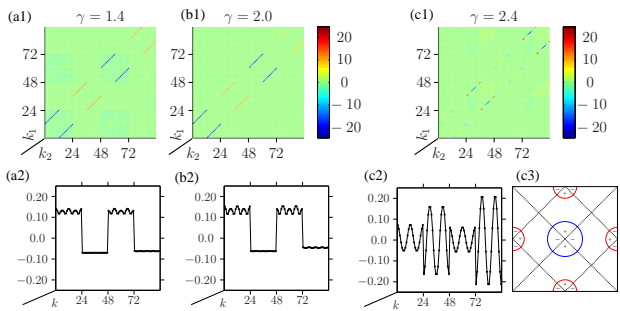}
  \end{minipage}
  \caption{(Color online) (a1)-(c1) plot of the vertex function with
    inter-band interaction $g_1=0.1$, $g_2=0.25$, $g_3=0.3$,
    $g_4=\gamma g_3$, $\gamma=2.4$, $b=1.1$, $x=-0.18$, and $\gamma =
    1.4, 2.0, 2.4$. For (a1) and (b2) we observe an ordinary $s^{\pm}$
    instability with small gap variation as shown in the form factor
    (a2) and (b2). In (c1), we observe a leading signal on the hole
    pockets originating from a $d$-wave pairing, where the nodes can
    be seen in the form factor (c2). (c3) gives a visualization of the
    extended $d$-wave state in the unfolded Brillouin zone. The bold
    dashed lines building the rhombohedron centered around the
    $\Gamma$ points correspond to $\cos k_x + \cos k_y$ yielding the
    sign change on the electron pockets around $(\pm \pi, 0)$ and $(0,
    \pm \pi)$. The dashed lines along the diagonals of the BZ
    correspond to the d-wave function $\cos k_x - \cos k_y$ labelling
    the nodal points on the hole pockets. Upon folding, we observe
    that the gaps on the line $\Gamma \rightarrow M$ have opposite
    signs.}
  \label{fig:hole}
\vspace{-10pt}
\end{figure}
We now consider hole doping to the system, and choose the same
interaction parameters $g_1=0.1$, $g_2=0.25$, $g_3=0.3$. For small
anisotropy $b$, the system likewise develops a rather constant
$s^{\pm}$ instability, where $T_c$ decreases with $\gamma$. For
considerable values of $b$ and the ratio $\gamma$
as for electron doping, we find that the hole doped scenario still
favors a constant $s^{\pm}$ instability (Fig.~\ref{fig:hole}). Unlike
in the electron-doped case, increasing $\gamma$ does not immediately
induce gap anisotropy on the electron pockets: the only result
(mostly) is a decreasing gap on the hole Fermi surfaces.  This
behavior can be explained due to the small electron FS size: for
hole-doping, the hole FS's dominate the behavior of the system. The
$A_{1g}$ term induces anisotropy mostly on the electron FS's, which at
hole doping are rather small. On the hole pockets, the $A_{1g}$ term
increases the size of the constant part of the interaction $g_3$,
which favors an $s^{\pm}$. This, coupled to the fact that the electron
FS's play a rather secondary role with respect to their hole
counterparts, renders the symmetry to be $s^{\pm}$ for moderate
$\gamma$. Increased $\gamma$ has the effect of reducing the scale of
the superconducting instability. Regarding the gap anisotropy, while
in the electron doped-case, increasing $\gamma$ resulted in nodal
electron FS's (while keeping the hole FS's isotropic nodeless), for hole
doping, the situation is completely different. Beyond a critical
$\gamma$, when the $s^{\pm}$ gaps have vanished, the system exhibits a
phase transition with nodal superconductivity on \emph{both} hole and
electron FS's. For $\gamma \gtrsim 2.4$, at comparably low transition
temperature, we find that the leading instability becomes a $d$-wave
intra-band Cooper pairing on the hole pockets. This is plausible from
the FS topology upon doping as shown in Fig.~\ref{fig:band}b: The hole
FS's around $\Gamma$ grow significantly upon hole doping. For dominant
intra-band repulsion $g_4$, the favorable ordering becomes a $d$-wave
Cooper pairing on the hole pockets, as shown in
Fig.~\ref{fig:hole}. 
\begin{figure}[t]

  \begin{minipage}[c]{0.95\linewidth}
    \includegraphics[width=\linewidth]{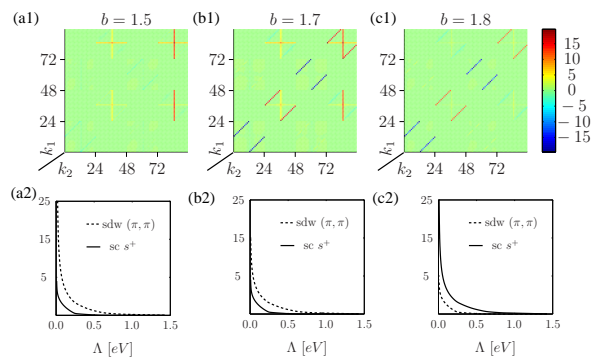}
  \end{minipage}
  \caption{(Color online)(a1)-(c1) Plot of the vertex function with
    interaction couplings
    $g_1=0.1$,$g_2=0.25$,$g_3=0.3$,$g_4=0.85$,$x=0.0$ and $b$ varying
    from left to right as $1.5$, $1.7$, and $1.8$. Upon increasing
    $b$, the previously subleading $s^\pm$ Cooper instability becomes
    more dominant and overcomes the SDW instability at $b \approx
    1.7$. (a2)-c(2) $s^\pm$ and SDW susceptibilities plotted for the
    different parameter settings. For $b=1.5$, one observes the SDW
    susceptibility (dashed line) to be dominant w. r. t. to the
    $s^\pm$ susceptibility. At $b=1.7$, the divergence scales are
    nearly equal, and $s^{\pm}$ becomes dominant for $b=1.8$.}
  \label{fig:b}
\vspace{-10pt}
\end{figure}
In terms of the hole FS only, this is a situation similar to the
cuprate superconductors. We also find a subdominant $d$-wave type
signal on the electron pockets, which relates to the $d$-wave signal
on the hole pockets shifted by $\pi$ so that there is still an overall
sign change in the gaps going from the hole pocket to the electron
pocket in the unfolded BZ; in this way, the system minimizes both
intra and inter-band repulsion: this is the $d$-wave equivalent of the
$s^\pm$ instability, which one may denote extended $d^{\pm}$-wave
(Fig.~\ref{fig:hole}c). This gap symmetry represents another,
time-reversal symmetric way to cope with the frustration pointed out
in Ref. \cite{lee-09prl217002}.

While interesting in its own right, the phase diagram on the
hole-doped side does not allow for a nodal electron FS \emph{while}
keeping the hole FS fully gapped. If this latter gap structure turns
out to be experimentally correct, we conclude that an $A_{1g}$
symmetry term alone is not sufficient to make the electron FS nodal
while keeping the hole FS fully gapped at both electron and hole
doping.

{\it Interplay of SDW and $s^{\pm}$ at half filling---} For all
parameter settings discussed in the previous sections, we always find
a leading SDW instability as we approach half filling. This is because
the SDW instability benefits enormously from the increased nesting of
hole and electron pockets. However, we observe that upon further
increasing $b$, the $s^{\pm}$ instability can overcome SDW and becomes
the leading instability even at half filling.  This is illustrated in
Fig.~\ref{fig:b}. There, we find that upon changing $b$ from $1.5$ to
$1.8$, the $s^{\pm}$ instability becomes dominant. The $b_c$ where
this occurs increases with $\gamma$. Interestingly, the critical scale
is not suppressed between the two regimes which in a less approximate
treatment would be most likely separated by a first order
transition. This means that a slight change of the system parameters
can turn the system from a high-scale SDW into a $s^\pm$ state with
comparable pairing scale - another possibility for a marked
material-dependence of the phase diagrams of different pnictides,
which may relate to recent measurements reported on
in~\cite{zhu-09cm09041732}.

\begin{figure}[t]

  \begin{minipage}[c]{0.95\linewidth}
    \includegraphics[width=\linewidth]{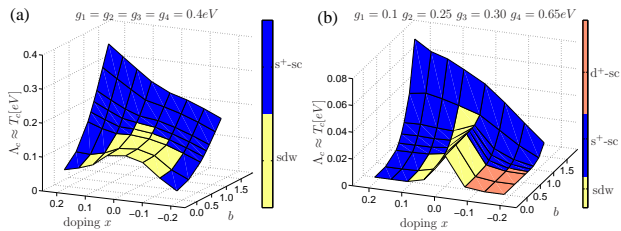}
  \end{minipage}
  \caption{(Color online). Phase diagrams for different interaction
    settings, (a) $g_1=g_2=g_3=g_4=0.4$ and (b) $g_1=0.1, g_2=0.25,
    g_3=0.3, g_4=\gamma g_3=0.65$. The three axes are given by the
    filling factor $x$, momentum anisotropy $b$ and the critical
    temperature $T_c$. (a) In a dome around half filling, the SDW
    instability is leading, until it is overcome by the
    superconducting instability at higher $b$ and electron (hole)
    doping. $T_c$ increases with $b$, in particular on the electron
    doped side. (b) $\gamma$ is increased. The SDW dome around half
    filling shrinks in doping width but increases on the anisotropy
    line as the superconducting $T_c$ is decreased due to $g_4$. For
    hole doping at comparably small amount of anisotropy $b$, we
    observe nodal $d^{\pm}$. $s^{\pm}$ gap variation on the electron
    doped side is enhanced, while true nodal extended s-wave is
    successively observed on the electron pockets only for even larger
    $\gamma \approx 3$ (see Fig.~\ref{fig:p2}).}
  \label{fig:p}
\vspace{-10pt}
\end{figure}

{\it{Full Phase Diagram---}} For various interactions settings, we
computed the complete phase diagram for different fillings and
anisotropy $b$, four representatives of which are shown with
increasing $\gamma$ in Fig.~\ref{fig:p} and Fig.~\ref{fig:p2}. We
observe that for comparably small $\gamma$, the phase diagram looks
rather uniform and only contains SDW and $s^{\pm}$ with moderate gap
variation as leading instabilities. For larger $\gamma$ and generally
reduced $T_c$ due to strong intra-band repulsion, the phase diagram
shows a more complex structure including the extended d-wave
instability on the hole doped side and the increasingly pronounced
$s^{\pm}$ gap variation for the electron doped side, which finally
yields a small parameter window of nodal $s^{\pm}$ for very large
$\gamma$ (Fig.~\ref{fig:p2}b). Thus, we find smoothly connected
instabilities at comparable scales, pointing to competing
orders. This, however, suggests a potentially high sensitivity to
material-specific parameters.

{\it Conclusion---} We studied the effect of certain momentum
dependence of band interactions for a 4-band model of the pnictides by
adding an $A_{1g}$-symmetric term to the pair-hopping interactions,
while keeping all other band interactions constant.  While most of our
data shows, in agreement with previous results, an $s^{\pm}$ wave
superconducting instability, on the electron doped side, we find that
increasing intra-band repulsion enhances the gap anisotropy on the
electron pockets, which may ultimately, but not readily, lead to a
true nodal electron FS gap.  For the hole-doped regime, however, the
gap anisotropy remains rather small even on the electron FS upon
increasing the interaction anisotropy, until a critical value of the
intra-band repulsion beyond which the system favors an interesting
state with intra-band $d$-wave cooper pairing on the hole pockets and
a reminiscent extended and sign-reversed $d$-wave signal on the
electron pockets. This represents a novel way to reduce the repulsion
within and between the Fermi surface pockets.  However, in the range
of our model, we are unable to find a regime which satisfies the
conditions of having gapped hole FS and nodal electron FS for
\emph{both} electron and hole doping. Finally we demonstrated that
even in the undoped state, small variations of the interaction
parameters can turn the SDW ground state into a superconducting state
with relatively high pairing scale.  Taken together, the outcome of
our studies is twofold. One possibility is that the bare interactions
have little anisotropy around the Fermi surfaces and the main
repulsive nesting is between electron and hole pockets, then the phase
diagrams should feature fully-gapped $s^\pm$ superconducting
states. Alternatively, if the experiments confirm gap nodes, this
would indicate a strong anisotropy of the bare interaction and a
stronger role of other repulsive interaction, e.g., between the
electron pockets. In this case, electron- and hole-doping may well
lead to different pairing states, and even undoped superconductivity
could be obtained for relatively mild parameter changes. One can ask
how much these findings depend on the type of anisotropy of the bare
interaction. In our understanding any pronounced wave
vector-dependence of the interaction will very likely cause
differences between electron and hole doping, as the relevant hole- or
electron Fermi surfaces are located in different regions of the
Brillouin zone.  Likewise, for the undoped system, a strong
interaction anisotropy generally enhances pairing states competing
with the SDW order such that material-specific parameter differences
might lead to observable consequences even in the undoped state. The
trends observed in our study should hold for any specific choice of
anisotropy in general.


\begin{acknowledgments}
We thank M. Eschrig, W. Hanke, D.-H. Lee, F. Wang, P. W\"olfle, and
  S. C. Zhang for discussions. CH and CP acknowledge support from
  DFG-FOR538 and FOR723, and BAB from the
  Alfred P. Sloan Foundation.
\end{acknowledgments}


\end{document}